\documentclass{emulateapj}   
\usepackage{apjfonts}
\submitted{ApJ Supp., accepted}

\def\ltsima{$\; \buildrel < \over \sim \;$}
\def\simlt{\lower.5ex\hbox{\ltsima}} 
\def\gtsima{$\; \buildrel > \over \sim \;$}
\def\simgt{\lower.5ex\hbox{\gtsima}} 
\def\arcsec{\hbox{$^{\prime\prime}$}}

\def\X{\textsf{X}}

\shorttitle{\X-shaped Radio Sources: II. Redshifts}
\shortauthors{Cheung et al.}

\begin{document}

\title{FIRST ``Winged'' and \X-shaped Radio Source Candidates: II. New 
Redshifts}

\author{C.~C. Cheung\altaffilmark{1,2,3}, 
Stephen~E. Healey\altaffilmark{3,4},
Hermine Landt\altaffilmark{5,6}, 
Gijs Verdoes Kleijn\altaffilmark{7}, 
Andr\'es Jord\'an\altaffilmark{5,8,9}}

\altaffiltext{1}{NASA Goddard Space Flight Center, Code 661, Greenbelt, MD 
20771, USA; Teddy.Cheung@nasa.gov.}

\altaffiltext{2}{Jansky Postdoctoral Fellow of the National Radio
Astronomy Observatory, USA.}

\altaffiltext{3}{Kavli Institute for Particle Astrophysics and 
Cosmology, Stanford University, Stanford, CA 94305, USA.}

\altaffiltext{4}{Department of Physics, Stanford University, Stanford, CA
94305, USA.}

\altaffiltext{5}{Harvard-Smithsonian Center for Astrophysics, 60 Garden St.,
Cambridge, MA 02138, USA.}

\altaffiltext{6}{Current address: School of Physics, University of 
Melbourne, Parkville, VIC 3010, Australia.}

\altaffiltext{7}{Kapteyn Astronomical Institute, Groningen, 9700 AV, The
Netherlands.}

\altaffiltext{8}{European Southern Observatory, 
Karl-Schwarzschild-Stra{\ss}e 2, 85748 Garching bei M\"unchen, Germany.}

\altaffiltext{9}{Current address: Departamento de Astronom\'{\i}a y 
Astrof\'{\i}sica, Pontificia Universidad Cat\'olica de Chile, Casilla
306, Santiago 22, Chile.}

\begin{abstract}

We report optical spectroscopic observations of \X-shaped radio sources 
with the Hobby-Eberly Telescope and Multiple-Mirror Telescope, focused 
on the sample of candidates from the FIRST survey presented in Paper I 
(Cheung 2007). A total of 27 redshifts were successfully obtained, 21 
of which are new, including that of a newly identified candidate source 
of this type which is presented here.  With these observations, the 
sample of candidates from Paper I is over 50$\%$ spectroscopically 
identified.  Two new broad emission-lined \X-shaped radio sources are 
revealed, while no emission lines were detected in about one third of 
the observed sources; a detailed study of the line properties is 
deferred to a future paper. Finally, to explore their relation to the 
Fanaroff-Riley division, the radio luminosities and host galaxy 
absolute magnitudes of a spectroscopically identified sample of 50 
\X-shaped radio galaxies are calculated to determine their placement in 
the Owen-Ledlow plane.

\end{abstract}


\section{Introduction}\label{section-intro}

The majority of double-lobed radio galaxies can be classified into one 
of two morphological classes: edge-dimmed and edge-brightened, 
corresponding to the \citet{fan74} type I and II sources, respectively.  
Some of the most unusual radio galaxies are those with an additional 
pair of extended low surface brightness `wings', making an overall 
\X-shaped appearance. One popular model posits that these are recently 
merged systems and that the wings are the inactive lobes marking the 
pre-merger axis of the central supermassive black hole/accretion disk 
system \citep[see, e.g.,][]{rot01,mer02,kom06}. However, this 
and other models \citep[e.g., buoyant expansion,][]{lea84,wor95} are not 
well tested.

Generally, these interpretations are hampered by the fact that a small 
number of \X-shaped radio sources are known \citep[a recent census has 
it at 16 bona fide examples;][]{che07} and an even smaller subset have 
been studied in detail. In an attempt to remedy this deficiency, 
\citet[][hereafter Paper~I]{che07} compiled a sample of 100 new 
candidate sources of this type selected from radio maps from the 
VLA-FIRST survey \citep{first}. One of the first steps in a systematic 
follow-up multi-wavelength study of this large sample is to obtain 
spectroscopic redshifts and identifications.  This is necessary to study 
their demographics, environments, etc., enabling comparisons with 
radio galaxies without wings \citep[cf.,][]{ulr96,zir97}. For instance, 
\X-shaped radio sources are known to have radio luminosities near the 
Fanaroff-Riley (FR) type I/II divide \citep{lea92,den02}, and their 
morphologies suggest that these may be the long sought-after transition 
sources (Paper I). To evaluate this interesting scenario, an important 
first step is to place the \X-shaped sources in the Owen-Ledlow 
\citep{owe93,led96} plane (radio luminosity versus host galaxy absolute 
magnitude).

As part of our \X-shaped source study, we have obtained optical 
spectroscopic observations of 26 sources included in Paper~I, and one 
additional new example (presented here). The redshifts, optical spectra, 
and some of the basic results from these observations are discussed. An 
Owen-Ledlow plot for a 50-source sample of spectroscopically identified 
\X-shaped radio galaxies and best candidates from Paper~I is constructed 
and this is discussed in relation to the Fanaroff-Riley division. A more 
detailed study of the spectra will be presented in a future paper in 
this series. Throughout, we assume $h=H_{0}$/(100 km s$^{-1}$ 
Mpc$^{-1}$)=0.7, $\Omega_{\rm M}=0.3$ and $\Omega_{\rm \Lambda}=0.7$, as 
in Paper~I.

\section{Observations and Analysis}\label{sec:obs}

This paper presents observations with the 9.2 m Hobby-Eberly Telescope 
(HET) at McDonald Observatory in 2006 and 2007 under queue observing 
mode and with the 6.5 m Multiple-Mirror Telescope (MMT) at Mt.\ Hopkins 
Observatory in a two-night run in Feb.\ 2007.  An observation of one 
object (J0702+5002) was obtained earlier with the 2.7 m Harlan J.\ Smith 
Telescope (HJST) at McDonald Observatory. A summary of the observations 
is presented in Table~\ref{table-1}.

On the HET, we used the Marcario Low-Resolution Spectrograph 
\citep[LRS;][]{hil98}, with grism G1, a $2\arcsec$ slit, and a Schott 
GG385 long-pass filter.  This provided broad wavelength coverage (from 
$\sim$4000--9200 \AA) and a resolution of $\mathcal{R} \approx 500$. 
Typically, we obtained $2\times300$~s exposures for the brightest 
targets and $2\times600$~s exposures for fainter ones. Eleven total 
spectra from the HET are presented here.

The MMT observations utilized the blue channel spectrograph 
($\mathcal{R} \approx 300$) with a $1\arcsec$ slit and wavelength range 
$\sim$3200--8400 \AA.  The slit was rotated to the parallactic angle for 
all observations. To mitigate against second-order effects, we used the 
L-42 filter on the second night of the observing run. Sky conditions 
were clear early on with increasing clouds toward the end of each night 
resulting in several hours where the telescope was idle. The seeing was 
consistently at the sub-arcsecond level for all 15 successfully obtained 
spectra.

The observation of J0702+5002, a candidate \X-shaped radio galaxy 
identified early on in the work leading up to Paper~I, was obtained by 
S.~E.~H.\ in 2005 with the 2.7 m HJST, using the Imaging Grism 
Instrument (IGI) and the 6000 \AA\ VPH grism.  The redshift was 
reported in Paper~I, and the spectrum is provided here.

Our targets were selected from the list of 100 FIRST \X-shaped radio 
source candidates presented in Paper~I (Table 2 therein). We created a 
prioritized list of targets consisting of candidates with radio 
morphologies judged most likely to be bona fide \X-shaped sources from 
their VLA-FIRST maps as well as those few known \X-shaped radio sources 
without known redshifts (Paper~I, Table~1 therein). To fill in 
scheduling gaps during the MMT run, we additionally observed several 
objects from the known sample where we did not have access to the 
optical spectra in digital format either from SDSS \citep{sdss} or 
privately available from other researchers. Lastly, one of us (C.~C.~C.) 
is compiling a new set of candidate \X-shaped radio sources as a 
follow-up study to Paper~I; a MMT spectrum of one of these 
was obtained, and the new identification is presented in \S~\ref{sec:new}.

All spectroscopic data were analyzed with IRAF\footnote{IRAF is 
distributed by the National Optical Astronomy Observatories, which are 
operated by the Association of Universities for Research in Astronomy, 
Inc., under cooperative agreement with the National Science Foundation.} 
using standard routines, including calibration of the absolute fluxes by 
comparison with standard stars and removal of telluric absorption 
features. Redshifts were determined from the narrow emission lines [O 
II] $\lambda$3727 and/or [O III] $\lambda$5007 for the majority of our 
sources (19/27 objects). In about one-third of our sources (8/27 
objects), no emission lines were detected, and we derived the redshift 
from (at least two of) the absorption features of the host galaxy 
(typically Ca II, G band, Mg Ib, and NaD).  We regard the redshifts of 
two sources, J1135--0737 and J1434+5906, as tentative because of the 
relatively low S/N ratio ($\sim$5) of their spectra.

\section{Results}

\begin{figure*}
\epsscale{1.0}
\plotone{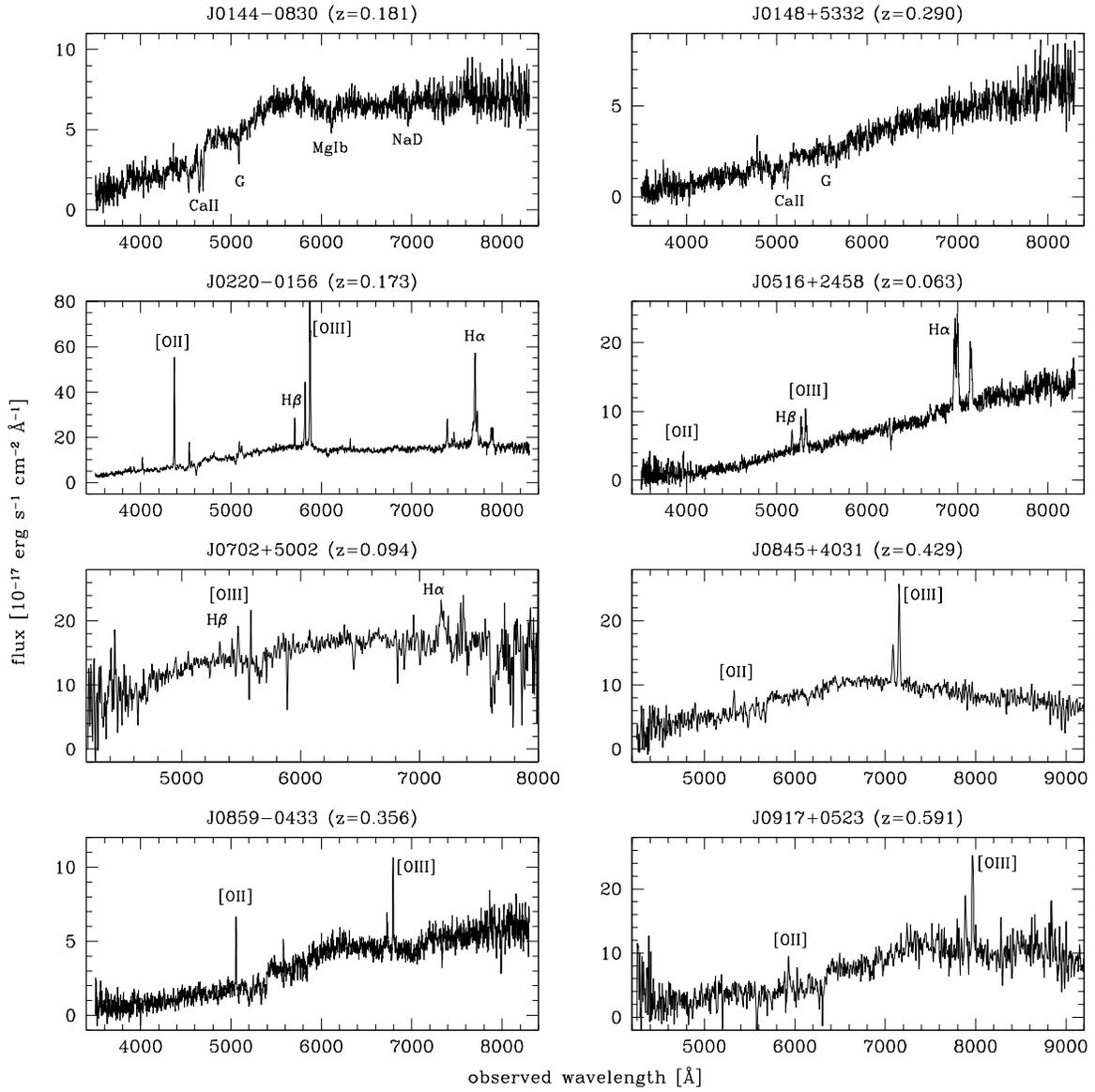}
\figcaption[f1.ps]{\label{figure-1}
Optical spectra of the targets presented in order of increasing right
ascension.}
\end{figure*}

The spectra are presented in Figure~\ref{figure-1}, and the redshift 
measurements are reported in Table~\ref{table-1}.  New spectroscopic 
identifications and redshifts have been successfully obtained for 21 
sources, and new spectra were obtained for 6 sources with known 
redshifts (\S~\ref{sec:known}). In 2 of the 21 cases in the former 
category, new redshifts are measured where only previous estimates were 
available (\S~\ref{sec:est}).

Since the publication of Paper~I, a spectroscopic redshift was found 
for the candidate ($\#$78 in Paper~I) \X-shaped radio source 
J1433+0037 \citep[$z=0.5031$;][]{can06}. Also, in \S~\ref{sec:miss}, 
we describe an optical misidentification of a candidate from Paper~I 
revealed by a new MMT spectrum. We note finally that other radio 
sources with \X-shaped or `winged' type morphologies are being 
identified \citep[e.g.,][]{lan08,sar08}. One new \X-shaped radio 
source, FIRST J1018+2914, discovered as part of a follow-up study to 
Paper~I, is described in \S~\ref{sec:new}.

\subsection{Spectra of Objects with Previously Reported 
Redshifts}\label{sec:known}

During the MMT run, we observed four well-known \X-shaped radio galaxies 
with previously determined redshifts: 3C~52 \citep{spi85}, 3C~63 
\citep{smi80}, 3C~136.1 \citep{smi76}, and J1101+1640 = Abell 1145 
\citep{owe95,owe97}. We additionally observed two candidate \X-shaped 
radio galaxies with known redshifts, J0941$-$0143 \citep{spi79} and 
J1614+2817 \citep{mil01}, which correspond to Catalog entries $\#$27 and 
$\#$94 in Paper I, respectively. The new spectra tend to be of higher 
quality than the previously obtained ones and the redshifts we measured 
are consistent ($\Delta z < 0.005$) with the published values for all 6 
sources.

We also observed J1357+4807, which was identified in Paper~I as an 
\X-shaped radio source from a published map but previously was without a 
spectroscopic identification \citep{leh01}. Its redshift of $z=0.383$ 
was reported in Paper~I, and the spectrum is now presented here. 
Similarly, in the case of the FIRST \X-shaped radio source candidate 
J0702+5002 (Catalog $\#$14 in Paper~I), we reported $z=0.094$ in 
Paper~I, and the spectrum is presented here.


\subsection{New Redshifts for Galaxies with Previous Estimates}\label{sec:est}

We obtained new spectra for two objects previously with only redshift 
estimates.  J1210$-$0341 (Catalog $\#$50 in Paper~I) has a newly 
measured redshift ($z=0.178$) while the previous estimate, quoted in 
Paper~I, was a photometric redshift of $z=0.26$ from \citet{mac99}. Our 
redshift of $z=0.358$ for J1253+3435 (TONS12$\_$301, Catalog $\#$59) is 
based on the detected emission lines from [O III] $\lambda\lambda$5007, 
4959 and [O II] $\lambda$3727 in our MMT spectrum. The earlier estimate 
of $z=0.034$ by \citet{bra05} was based on attributing the [O III] 
$\lambda$5007 emission line observed at 6800 \AA\ in their lower S/N 
spectrum to H$\alpha$.

\section{\X-shaped Radio Galaxies and the Fanaroff-Riley Division}\label{sec:fr}

In Table~\ref{table-2}, we define a sample of 50 spectroscopically 
identified \X-shaped radio galaxies from the known list and the best 
candidates from Paper~I. Of this radio galaxy\footnote{Based on their 
optical spectra, in particular the absence of broad emission lines, the 
optical light is most likely dominated by the host galaxy and little 
contaminated by relativistically beamed emission from a jet. The broad 
emission-lined sources are not included at this time because of the 
possible contamination of the nuclear emission.} sample, 15 known 
sources and 16 VLA-FIRST candidates had redshifts available from the 
literature; this paper is responsible for the inclusion of the remaining 
19 sources with our newly determined redshifts. For this sample, we can 
determine the source distances, thus allowing estimates of the absolute 
optical magnitudes of their host galaxies and the (monochromatic) radio 
luminosities.

We find that the average absolute $R$-band magnitude of the \X-shaped 
radio galaxy sample is $-$23.2 (1$\sigma = 0.8$; median = $-$23.1) 
which is entirely consistent with values for `normal' radio galaxies. 
Specifically, \citet{gov00} found an average of $M_{\rm R} = -23.33 \pm 
0.69$ (converting to our adopted cosmology) for their sample which 
includes both FR~I and FR~II sources, and \citet{mcl04} found an 
average $M_{\rm R} = -23.20 \pm 0.09$ (median = $-$23.25) in their 
sample of powerful $z\sim 0.5$ radio galaxies.

As found in Paper~I, we confirm that the average 1.4 GHz radio 
luminosity of our sample is close to the Fanaroff-Riley division 
(average log $L_{\rm 1.4}$ [W Hz$^{-1}$]= 25.79, 1$\sigma = 0.69$). This 
is an expected result since \X-shaped radio galaxies have long been 
known to have typical radio luminosities near the FR~I/II division 
\citep{lea92,den02}.

To further explore the relationship of \X-shaped radio galaxies with the 
Fanaroff-Riley division, in Figure~\ref{figure-lm}, we plot the 1.4 GHz radio 
luminosity vs. parent host galaxy absolute $R$-magnitude diagram for this 
sample. Owen \& Ledlow \citep{owe93,led96} found a clear dividing line between 
FR~I and FR~II radio sources in this diagnostic plane, and this line is drawn 
in Figure~\ref{figure-lm}. As anticipated from their average radio 
luminosities, the \X-shaped sources straddle the Owen-Ledlow FR~I/II dividing 
line, but there is a large dispersion. For the sample of known \X-shaped radio 
galaxies (marked with an `X' in Figure~\ref{figure-lm}), the objects are 
neatly separated by their Fanaroff-Riley type morphology with the one clear 
FR~I radio galaxy \citep[NGC326 = X02;][]{eke78,mur01} clearly below this 
line. The only potential exception is that of 3C~315 (X14), whose radio 
morphology on the scale of hundreds of kpc is more typical of an FR~I 
\citep{lea84,lal07} but which lies in the FR~II regime of the Owen-Ledlow plot 
(Figure~\ref{figure-lm}). However, its central radio component is remarkable 
in that it is resolved into a few kpc-scale FR~II-like double 
\citep{dek00,sar08}, although this feature is responsible for only a small 
fraction of the source total luminosity \citep[see the maps 
in][]{lea84,lal07}. A few of the FIRST candidates from Paper~I also extend 
into the FR~I portion of the Owen-Ledlow plot. New VLA imaging of the sample 
(already obtained for many of them) will allow us to classify their FR type to 
explore this relationship further in a future paper. At present, the 
connection between the \X-shaped source phenomenon and the Fanaroff-Riley 
division remains unclear.

\section{Discussion and Summary}

New optical spectroscopic observations have been obtained for 27 
candidate and known \X-shaped radio sources. Six had previously known 
redshifts (\S~\ref{sec:known}), and 21 are newly determined redshifts 
(19 candidates from Paper~I, one previously known, and one new \X-shaped 
source). With only 34/100 \X-shaped radio source candidates presented in 
Paper~I with redshifts previously available from the literature, this 
now brings the sample of candidates over 50$\%$ identified. More 
importantly, since our spectroscopic targets were selected as the most 
promising \X-shaped source candidates based on their morphologies in the 
FIRST maps, a large fraction of the most interesting sources are now 
spectroscopically identified (see \S~\ref{sec:fr} and 
Table~\ref{table-2}).

The sources with redshifts determined in this work are predominantly 
fainter optically than the 34 \X-shaped source candidates with 
previously available redshifts in Paper~I (see Figure~\ref{figure-rz}). 
As expected, our targets are found to be typically more distant than 
previously known examples ($z<0.4$; Paper~I).  With 8 sources newly 
determined to have redshifts of $z>0.5$ (Figure~\ref{figure-rz}), this 
extends our census of \X-shaped sources to higher redshifts. It is worth 
mentioning that half of the redshifts taken from the literature were 
obtained from the SDSS \citep{sdss}, and these were fairly bright ($r$ 
\simlt 18--19) targets; our targets are typically fainter and thus will 
likely not be observed by SDSS in the future.

Roughly a third of the observed sources have no emission lines detected 
and the remaining are predominantly strong narrow emission-line objects. 
The obvious exceptions are J1342+2547 and J1406+0647 (Catalog $\#$67 and 
$\#$73, respectively), which also have strong broad emission lines 
indicative of quasars.  Broad-line emission from these two targets was 
indeed anticipated in Paper I based on their bluer optical colors ($g - 
r = 0.4$ and 0.0, respectively). This adds to the 2 known and 4 
candidate broad lined \X-shaped sources presented in Paper I. Such broad 
line objects seem to be relatively rare amongst \X-shaped radio sources, 
occurring in only $\sim$10$\%$ of the sample, and these are two 
excellent examples. The overall rate of detected emission line objects/types 
here is consistent with that of normal radio galaxies 
\citep[e.g.,][]{tad93}.

It is important to identify more broad emission-lined objects as 
they may provide valuable clues as to the possible formation scenarios 
for \X-shaped radio morphologies. Specifically, the spectra can be used 
to constrain systematic velocity offsets between the broad and narrow 
lines, a possible observational signature of gravitational wave 
radiation recoil following a supermassive black hole binary merger 
\citep{bon07,kom08}. These, and other aspects of the spectra, will be 
explored in a future paper to search further for possible clues as to 
the origin of the unusual morphologies of these radio sources.


\acknowledgments
\begin{center}Acknowledgments\end{center}

C.~C.~C.\ was supported by the National Radio Astronomy Observatory 
(2004--2007) which is operated by Associated Universities, Inc.\ under a 
cooperative agreement with the National Science Foundation, and 
currently by an appointment to the NASA Postdoctoral Program at Goddard 
Space Flight Center, administered by Oak Ridge Associated Universities 
through a contract with NASA. S.~E.~H.\ is supported by the Stanford 
Linear Accelerator Center under DOE contract DE-AC03-76SF00515.

C.~C.~C.\ is grateful to the MMT staff, particularly Mike Alegria and 
Grant Williams for their assistance during the observing run, and 
Alessondra Springmann for her assistance during a clouded out MMT run in 
July 2006. The MMT is a facility operated jointly by the University of 
Arizona and the Smithsonian Institution.

The Hobby-Eberly Telescope is a joint project of the University of Texas 
at Austin, the Pennsylvania State University, Stanford University, 
Ludwig-Maximillians-Universit\"at M\"unchen, and 
Georg-August-Universit\"at G\"ottingen. The HET is named in honor of its 
principal benefactors, William P.\ Hobby and Robert E.\ Eberly.  The 
Marcario Low-Resolution Spectrograph is named for Mike Marcario of High 
Lonesome Optics, who fabricated several optics for the instrument but 
died before its completion; it is a joint project of the HET partnership 
and the Instituto de Astronom\'{\i}a de la Universidad Nacional 
Aut\'onoma de M\'exico.  This paper includes data taken with the Harlan 
J.\ Smith Telescope at the McDonald Observatory of the University of 
Texas at Austin.

This research has made use of NASA's Astrophysics Data System Abstract 
Service and the NASA/IPAC Extragalactic Database (NED) which is operated 
by the Jet Propulsion Laboratory, California Institute of Technology, 
under contract with the National Aeronautics and Space Administration.

{\it Facilities:} \facility{HET}, \facility{MMT}, \facility{VLA}

\appendix

\section{Optical Misidentification of J1258+3227}\label{sec:miss}

A MMT spectrum (single 5 minute exposure) of the bright ($r=17$) optical
object SDSS~J125832.87+322740.8 identified with the \X-shaped radio
source candidate J1258+3227 (Catalog $\#$60 in Paper~I) revealed it to
be a G-type star. In consideration of this result, the better choice for
an optical counterpart to the radio source is the fainter ($r=21.1$ mag,
$g-r=1.5$) galaxy SDSS J125833.29+322737.1, located $\sim$7\arcsec\ to
the southeast of the brighter star. The fainter optical source is closer
to the central region of the radio source and is not visible in the
shallower DSS image presented in Paper I (Figure~2 therein).

\section{A New \X-shaped Radio Source}\label{sec:new}

FIRST J1018+2914 (4C+29.38, B2~1015+29) was identified as a clear \X-shaped
radio source in its VLA-FIRST map (Figure~\ref{figure-new}) over the
course of follow-up work to Paper~I by one of us (C.~C.~C.). The radio
source is identified with SDSS J101827.09+291420.3, a $r=18.7$ mag
($g-r=1.4$) extended galaxy \citep{sdss}.  The radio source displays a
steep spectrum with flux densities at 0.365, 1.4, and 4.9 GHz of 1037
mJy \citep{dou96}, 444 mJy \citep{con98}, and 154 mJy \citep{gre91},
respectively.  With our new MMT spectrum, we measured $z=0.389$ for the
galaxy.


{}


\addtocounter{figure}{0}
\begin{figure*}
\epsscale{1.0}
\plotone{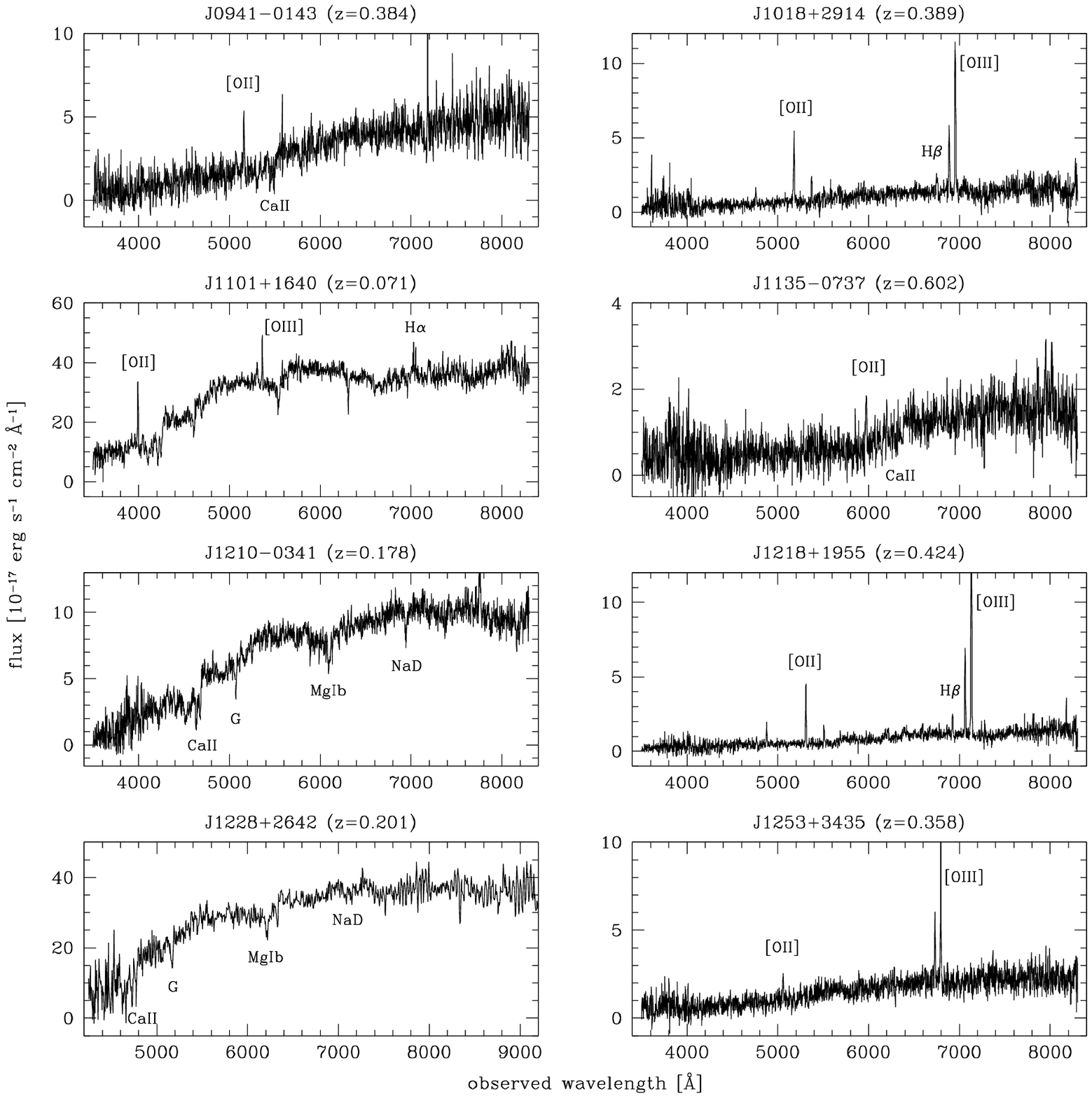}
\end{figure*}

\addtocounter{figure}{0}
\begin{figure*}
\epsscale{1.0}
\plotone{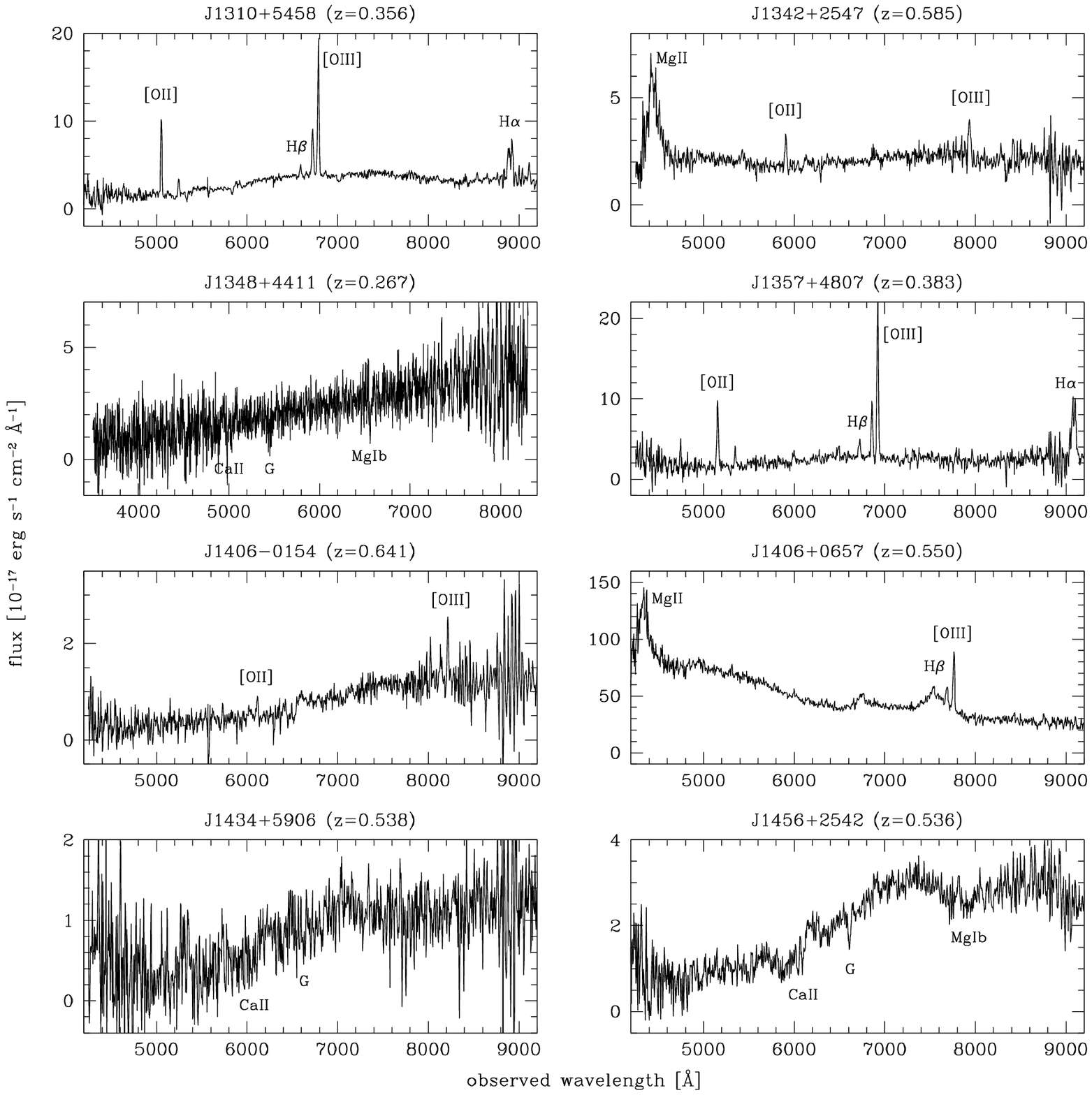}
\end{figure*}

\addtocounter{figure}{0}
\begin{figure*}
\epsscale{1.0}
\plotone{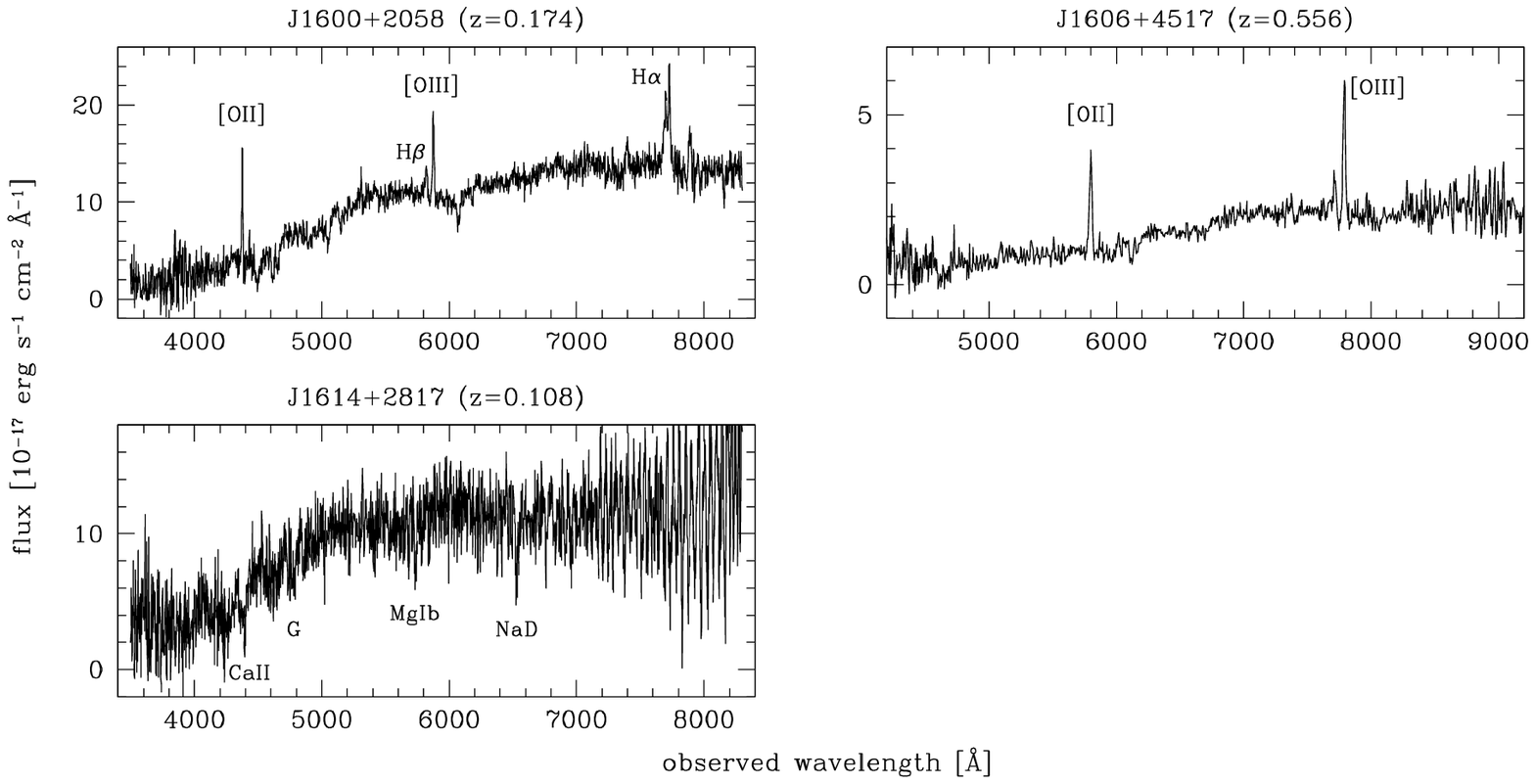}
\end{figure*}

\begin{figure}
\epsscale{0.5}
\plotone{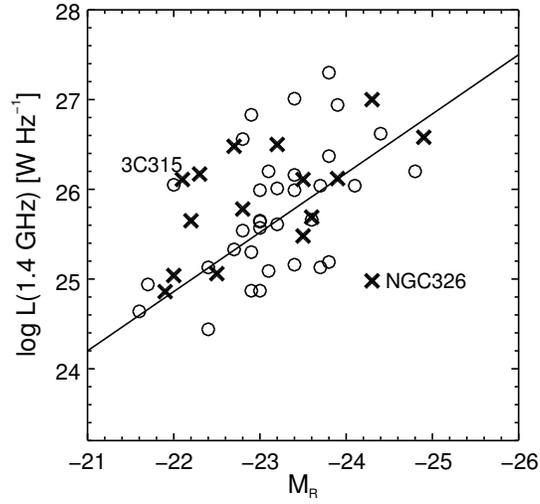}
\figcaption[f2.eps]{\label{figure-lm}
Radio luminosity at 1.4 GHz versus absolute optical magnitude in the
$R$-band for all known (crosses) and our best candidate (dots) \X-shaped
radio galaxy sample -- see Table~\ref{table-2}. The line is the observed
division in this plane between FR~I (below) and FR~II (above) radio
sources \citep{owe93,led96}, as parameterized by \citet{wol07} who
converted to integrated magnitudes and a modern cosmology --
it is {\emph not} a fit to the plotted
data. The points for two interesting known
sources described in \S~\ref{sec:fr} are marked.} \end{figure}

\begin{figure}
\epsscale{0.5}
\plotone{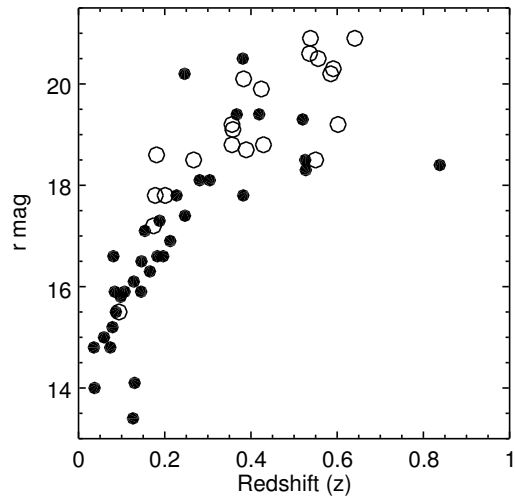}
\figcaption[f3.eps]{\label{figure-rz}
Optical $r$-band magnitude versus redshift distribution for all FIRST
\X-shaped radio source candidates from Paper~I (Table~2 therein)
previously with redshifts available (filled circles) and for the sources
with newly obtained redshifts presented in this paper (open circles).
Notice that the newly targeted sources tend to be fainter optical
objects and at systematically higher redshifts.} \end{figure}

\begin{figure}
\epsscale{0.5}
\plotone{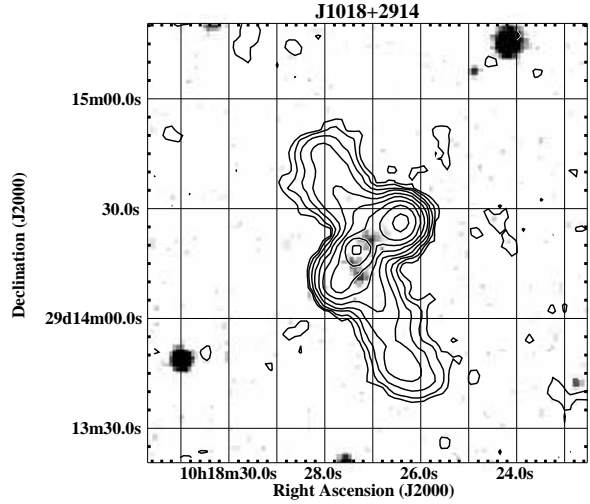}
\figcaption[f4.eps]{\label{figure-new}
Radio image at 1.4 GHz from the VLA-FIRST survey (contours) of the new
\X-shaped radio galaxy FIRST J1018+2914 described in \S~\ref{sec:new} overlaid
onto a DSS2 (red) optical image.} \end{figure}



\begin{deluxetable}{lccccclcc}
\tabletypesize{\scriptsize}
\tablecaption{Observation Log and Measured Redshifts\label{table-1}}
\tablewidth{0pt}
\tablehead{
\colhead{Cat.~$\#$\tablenotemark{a}} & \colhead{Name\tablenotemark{b}} & 
\colhead{Mag.\tablenotemark{c}} & \colhead{Telescope\tablenotemark{d}} & 
\colhead{Obs. Date} & \colhead{Exp.\ Time\tablenotemark{e}} & 
\colhead{$z$\tablenotemark{f}} & \colhead{Lines\tablenotemark{f}} & 
\colhead{Notes\tablenotemark{g}}
}
\startdata
9  &J0144$-$0830 & 18.6 &MMT  &2007 Feb 07  &1200 & 0.181 & abs& \\
X03&J0148$+$5332 & 17.5 &MMT  &2007 Feb 07  &1800 & 0.290 & abs& 1 \\
X04&J0220$-$0156 & 17.8 &MMT  &2007 Feb 07  &1200 & 0.173 & em & 1 \\
X05&J0516$+$2458 & 17.0 &MMT  &2007 Feb 08  &\phantom{1}600 & 0.063 & em & 1 \\
14 &J0702$+$5002 & 15.5 &HJST &2005 Oct 30  &\phantom{1}600 & 0.094 & em & 2\\
21 &J0845$+$4031 & 18.8 &HET  &2006 Apr 30  &\phantom{1}600 & 0.429 & em & \\
23 &J0859$-$0433 & 18.8 &MMT  &2007 Feb 07  &1200 & 0.356 & em &\\
25 &J0917$+$0523 & 20.3 &HET  &2006 May 01  &1200 & 0.591 & em &\\
27 &J0941$-$0143 & 17.8 &MMT  &2007 Feb 07  &\phantom{1}900 & 0.384 & em &1\\
   &J1018$+$2914 & 18.7 &MMT  &2007 Feb 08  &1500 & 0.389 & em &3\\
X10&J1101$+$1640 & 15.9 &MMT  &2007 Feb 07  &\phantom{1}600 & 0.071 & em & 1 \\
43 &J1135$-$0737 & 19.2 &MMT  &2007 Feb 09  &1800 & 0.602 & em & 4\\
50 &J1210$-$0341 & 17.8 &MMT  &2007 Feb 09  &1000 & 0.178 & abs & 5\\
53 &J1218$+$1955 & 19.9 &MMT  &2007 Feb 09  &3300 & 0.424 & em & \\
56 &J1228$+$2642 & 17.8 &HET  &2006 May 09  &1200 & 0.201 & abs & \\
59 &J1253$+$3435 & 19.1 &MMT  &2007 Feb 09  &2300 & 0.358 & em & 5\\
62 &J1310$+$5458 & 19.2 &HET  &2006 Jun 26  &1200 & 0.356 & em & \\
67 &J1342$+$2547 & 20.2 &HET  &2006 Apr 21  &1200 & 0.585 & em & \\
69 &J1348$+$4411 & 18.5 &MMT  &2007 Feb 08  &1800 & 0.267 & abs & \\
X13&J1357$+$4807 & 20.1 &HET  &2006 Apr 07  &\phantom{1}600 & 0.383 & em & 2 \\
72 &J1406$-$0154 & 20.9 &HET  &2006 May 23  &1200 & 0.641 & em & \\
73 &J1406$+$0657 & 18.5 &HET  &2007 May 05  &\phantom{1}600 & 0.550 & em & \\
79 &J1434$+$5906 & 20.9 &HET  &2006 Apr 09  &\phantom{1}600 & 0.538 & abs & 4\\
84 &J1456$+$2542 & 20.6 &HET  &2007 Apr 26  &1200 & 0.536 & abs&\\
90 &J1600$+$2058 & 17.2 &MMT  &2007 Feb 09  &\phantom{1}900 & 0.174 & em & \\
93 &J1606$+$4517 & 20.5 &HET  &2007 May 14  &1200 & 0.556 & em & \\
94 &J1614$+$2817 & 15.9 &MMT  &2007 Feb 08  &\phantom{1}600 & 0.108 & abs& 1\\
\enddata
\footnotesize
\tablenotetext{a}{Catalog number based on entry order in Table~1 
(prepended with X) and Table~2 of Paper~I \citep{che07} containing the 
lists of the known and candidate \X-shaped radio sources, respectively.}

\tablenotetext{b}{Object name based on J2000 coordinates.}

\tablenotetext{c}{Optical magnitudes predominantly from Paper I -- see 
summary in Table~\ref{table-2}.}

\tablenotetext{d}{HET = Hobby-Eberly Telescope, HJST = Harlan J.\ Smith 
Telescope, MMT = Multi-Mirror Telescope.}

\tablenotetext{e}{Total exposure time in seconds.}

\tablenotetext{f}{Measured redshifts determined from emission (em) or 
absorption (abs) features.}

\tablenotetext{g}{Notes on redshifts: 
(1) redshift previously known (see \S~\ref{sec:known}), 
(2) redshift reported in Paper I as described in \S~\ref{sec:known}, 
(3) new FIRST \X-shaped radio source candidate not presented in Paper I 
and described in \S~\ref{sec:new},
(4) tentative redshift due to relatively low S/N ($\sim$5) of the spectrum,
(5) new redshift superseding previous estimate (see \S~\ref{sec:est}).}

\end{deluxetable}

\begin{deluxetable}{lccccccc}
\tabletypesize{\scriptsize}
\tablecaption{Parameters of Spectroscopically Identified
\X-shaped Radio Galaxy Sample\label{table-2}}
\tablewidth{0pt}
\tablehead{
\colhead{Cat.~$\#$\tablenotemark{a}} & \colhead{Name\tablenotemark{b}} & 
\colhead{Mag.\tablenotemark{c}} & \colhead{$z$\tablenotemark{d}} & 
\colhead{$K$-Corr.\tablenotemark{e}} & \colhead{$A$\tablenotemark{f}} & 
\colhead{log $L_{\rm 1.4}$\tablenotemark{g}} & 
\colhead{$M_{\rm R}$\tablenotemark{h}}
}
\startdata
X01     &J0009+1244    &15.9*   &0.156  &0.17   &0.24   &26.12  &$-$23.9  \\
X02     &J0058+2651    &13.2*   &0.0477 &0.05   &0.23   &24.98  &$-$24.3  \\
X03     &J0148+5332    &17.5*   &0.2854 &0.36   &0.62   &27.00  &$-$24.3  \\
X04     &J0220$-$0156  &17.8*   &0.175  &0.20   &0.09   &26.48  &$-$22.7  \\
X05     &J0516+2458    &17.0*   &0.064  &0.07   &2.53   &25.48  &$-$23.5  \\
X06     &J0805+2409    &15.4    &0.0598 &0.06   &0.15   &25.65  &$-$22.2  \\
X07     &J0831+3219    &14.7    &0.0507 &0.05   &0.13   &25.06  &$-$22.5  \\
X08     &J0941+3944    &16.1    &0.1075 &0.12   &0.05   &25.78  &$-$22.8  \\
X09     &J1020+4831    &15.1    &0.052  &0.05   &0.03   &25.04  &$-$22.0  \\
X10     &J1101+1640    &15.9    &0.068  &0.08   &0.05   &24.86  &$-$21.9  \\
X13     &J1357+4807    &20.1    &0.383  &0.56   &0.04   &26.17  &$-$22.3  \\
X14     &J1513+2607    &16.9    &0.1083 &0.12   &0.16   &26.11  &$-$22.1  \\
X15     &J1824+7420    &16.1*   &0.256  &0.31   &0.17   &26.58  &$-$24.9  \\
X16     &J1952+0230    &14.1*   &0.059  &0.06   &0.50   &25.69  &$-$23.6  \\
X17     &J2123+2504    &16.4*   &0.1016 &0.11   &0.48   &26.50  &$-$23.2  \\
X18     &J2157+0037    &19.1    &0.3907 &0.58   &0.14   &26.11  &$-$23.5  \\
1       &J0001$-$0033  &17.4    &0.2469 &0.30   &0.10   &25.13  &$-$23.7  \\
5       &J0049+0059    &18.1    &0.3044 &0.40   &0.06   &25.66  &$-$23.6  \\
6       &J0113+0106    &18.1    &0.281  &0.35   &0.09   &25.99  &$-$23.4  \\
7       &J0115$-$0000  &20.5    &0.381  &0.56   &0.08   &26.05  &$-$22.0  \\
9       &J0144$-$0830  &18.6    &0.181  &0.20   &0.07   &24.64  &$-$21.6  \\
14      &J0702+5002    &15.5*   &0.094  &0.10   &0.19   &24.87  &$-$23.0  \\
17      &J0813+4347    &16.1    &0.1282 &0.14   &0.19   &25.16  &$-$23.4  \\
21      &J0845+4031    &18.8    &0.429  &0.69   &0.10   &26.04  &$-$24.1  \\
23      &J0859$-$0433  &18.8*   &0.356  &0.50   &0.06   &26.01  &$-$23.2  \\
25      &J0917+0523    &20.3    &0.591  &1.18   &0.11   &26.94  &$-$23.9  \\
26      &J0924+4233    &17.8    &0.2274 &0.27   &0.05   &25.65  &$-$23.0  \\
27      &J0941$-$0143  &17.8*   &0.382  &0.56   &0.08   &26.62  &$-$24.4  \\
30      &J1005+1154    &16.3    &0.1656 &0.19   &0.12   &25.19  &$-$23.8  \\
43      &J1135$-$0737  &19.2*   &0.602  &1.21   &0.09   &26.20  &$-$24.8  \\
44      &J1140+1057    &16.6    &0.0808 &0.09   &0.16   &24.94  &$-$21.7  \\
49      &J1207+3352    &15.2    &0.0788 &0.09   &0.04   &24.87  &$-$22.9  \\
50      &J1210$-$0341  &17.8    &0.178  &0.20   &0.09   &25.13  &$-$22.4  \\
53      &J1218+1955    &19.9    &0.424  &0.67   &0.07   &26.83  &$-$22.9  \\
56      &J1228+2642    &17.8    &0.201  &0.23   &0.06   &25.33  &$-$22.7  \\
59      &J1253+3435    &19.1    &0.358  &0.51   &0.04   &26.20  &$-$23.1  \\
61      &J1309$-$0012  &19.4    &0.419  &0.66   &0.07   &27.01  &$-$23.4  \\
62      &J1310+5458    &19.2    &0.356  &0.50   &0.05   &25.99  &$-$23.0  \\
64      &J1327$-$0203  &16.6    &0.1828 &0.20   &0.09   &26.04  &$-$23.7  \\
69      &J1348+4411    &18.5    &0.267  &0.34   &0.03   &25.54  &$-$22.8  \\
72      &J1406$-$0154  &20.9    &0.641  &1.34   &0.16   &27.30  &$-$23.8  \\
76      &J1424+2637    &14.0    &0.0372 &0.04   &0.05   &24.44  &$-$22.4  \\
79      &J1434+5906    &20.9    &0.538  &1.00   &0.03   &26.56  &$-$22.8  \\
81      &J1444+4147    &17.3    &0.188  &0.21   &0.04   &25.57  &$-$23.0  \\
84      &J1456+2542    &20.6    &0.536  &1.00   &0.11   &25.61  &$-$23.2  \\
90      &J1600+2058    &17.2    &0.174  &0.19   &0.19   &25.64  &$-$23.0  \\
92      &J1606+0000    &15.0    &0.059  &0.06   &0.51   &25.30  &$-$22.9  \\
93      &J1606+4517    &20.5    &0.556  &1.06   &0.04   &26.16  &$-$23.4  \\
94      &J1614+2817    &15.9    &0.1069 &0.12   &0.13   &25.09  &$-$23.1  \\
        &J1018+2914    &18.7    &0.389  &0.58   &0.07   &26.37  &$-$23.8  \\
\enddata
\footnotesize
\tablenotetext{a}{Catalog number as in Table~\ref{table-1}.}

\tablenotetext{b}{Object name based on J2000 coordinates.}

\tablenotetext{c}{The observed optical magnitudes are predominantly SDSS 
$r$-band values unless otherwise indicated with an asterisk. For the 
known \X-shaped radio sources (Cat.~$\#$ with an `X' prefix), SDSS 
measurements are available for $\sim$1/2 of the sample and are reported 
here (sources X06 to X14, and X18). For the remaining objects, less 
accurate (by as much as $\sim$0.5--1 mag) Cousins $R$-band magnitudes 
[X01 \citep{wol07}, X16 \citep{gov00}, and X03, X15, and X19 
\citep[\mbox{USNO-B1 catalog};][]{usno}], and $V$-band magnitudes [X02 
\citep{dev91}, X04 and X17 \citep{smi89}, and X05 \citep{spi85}] are 
provided. For the candidates (Cat.~$\#$ without an `X' prefix), the 
magnitudes are tabulated in Table~2 of Paper I, where $R$-band values 
from the \mbox{USNO-B1} catalog were utilized for Cat. sources $\#$ 14, 
23, 27, and 43.}

\tablenotetext{d}{Redshifts are gathered from Paper~I (Tables~1 and 2 
therein) and Table~\ref{table-1} of this paper and are listed here for 
convenience.}

\tablenotetext{e}{$K$-corrections adopted based on $R$-band values from 
\citet{fuk95} as tabulated in \citet{urr00}. In the few cases where 
$V$-band measurements were used (all $z < 0.2$), the additional 
uncertainty due to utilizing the $K$-correction at $R$-band is 
negligible in comparison to the uncertainty in the published 
magnitudes.}

\tablenotetext{f}{Extinction corrections adopting $R$- and $V$-band 
values based on the \citet{sch98} maps.}

\tablenotetext{g}{Logarithm of the total source radio luminosity at 1.4 
GHz. The radio fluxes used were tabulated in Paper~I (Tables~1 and 2 
therein).}

\tablenotetext{h}{Absolute $R$-band magnitude with $K$- and extinction 
corrections applied. Transformations assumed colors of $r-R = 0.25$ and 
$V-R = 0.61$ appropriate for giant ellipticals \citep{fuk95}.}

\end{deluxetable}

\end{document}